\documentclass[preprint,double-spaced,showpacs,amsmath,amssymb]{revtex4}
\usepackage{graphicx}
\usepackage{dcolumn}

\begin{document}

\newcommand{\beq}{\begin{equation}}
\newcommand{\eeq}{\end{equation}}
\newcommand{\bea}{\begin{eqnarray}}
\newcommand{\eea}{\end{eqnarray}}
\newcommand{\eps}{\varepsilon}
\newcommand{\Fs}{\mbox{\scriptsize F}}
\newcommand{\pr}{_{\perp}}
\newcommand{\bk}{{\bf k}}
\newcommand{\bs}{{\bf s}}
\newcommand{\bp}{{\bf p}}
\newcommand{\br}{{\bf r}}
\newcommand{\bR}{{\bf R}}
\newcommand{\lsim}{\stackrel{\scriptstyle <}{\phantom{}_{\sim}}}
\newcommand{\gsim}{\stackrel{\scriptstyle >}{\phantom{}_{\sim}}}
\newcommand{\Vef}{V^{p}_{\mbox{\scriptsize eff}}}

\newcommand{\VF}{{\mathcal V}^{\mbox{\scriptsize F}}_{\mbox{\scriptsize eff}}}

\title{ Spatial correlation properties of the anomalous density matrix\\ in a slab of
nuclear matter with realistic NN-forces}

\author{ S.\,S. Pankratov}
\email{pankratov@mbslab.kiae.ru}
\affiliation{Kurchatov Institute, 123182, Moscow, Russia}

\author{ M. Baldo}
\email{baldo@ct.infn.it} \affiliation{INFN, Sezione di Catania, 64
Via S.-Sofia, I-95125 Catania, Italy}

\author{ U. Lombardo}
\email{lombardo@lns.infn.it} \affiliation{INFN-LNS and University
of Catania, 44 Via S.-Sofia, I-95125 Catania, Italy}

\author{ E.\,E. Saperstein}
\email{saper@mbslab.kiae.ru}
\affiliation{Kurchatov Institute, 123182, Moscow, Russia}

\author{ M.\,V. Zverev}
\email{zverev@mbslab.kiae.ru} \affiliation{Kurchatov Institute,
123182, Moscow, Russia}

\date{\today}

\begin{abstract}
Spatial correlation characteristics of the anomalous density
matrix $\varkappa$ in a slab of nuclear matter with the Paris and
Argonne v$_{18}$ forces are calculated. A detailed comparison with
predictions of the effective Gogny force is made. It is found that
the two realistic forces lead to very close results which are
qualitatively similar to those for the Gogny force. At the same
time, the magnitude of $\varkappa$ for realistic forces is
essentially smaller than the one for the Gogny force. The
correlation characteristics  are practically independent of the
magnitude of $\varkappa$ and turn out to be quite close for the
three kinds of the force. In particular, all of them predict a
small value of the local correlation length at the surface of the
slab and a big one, inside. These results are in agreement with
those obtained recently by Pillet at al. for finite nuclei with
the Gogny force.

\end{abstract}

\pacs{21.30.Cb; 21.30.Fe; 21.60.De}

\maketitle

\section{Introduction}

The problem of surface nature of nuclear pairing has a long
history, see the review paper \cite{Rep} and Refs. therein. First,
it was formulated in terms of the effective pairing interaction
(EPI) entering the gap equation in a model space in which the
pairing problem is usually considered. Within the self-consistent
Finite Fermi System theory, the use of a natural density dependent
ansatz for the EPI has resulted in a strong attraction at the
nuclear surface, being rather small inside nuclei \cite{ZvSap}.
Similar conclusion was obtained in the {\it ab initio} calculation
of the EPI \cite{BLSZ1}. Later, the surface enhancement of the gap
function $\Delta(X)$ was found by solving the gap equation for the
complete Hilbert space in semi-infinite nuclear matter with the
realistic Paris force in \cite{BLSZ2} and  with the effective
Gogny force in \cite{FarSch}. Similar conclusions were obtained
for a nuclear slab in Ref. \cite{6auth} where the gap equation was
solved for both the types of NN force simultaneously. It was found
that, although there is a quntitative difference between
predictions of the two calculations, both of them show  a
pronounced maximum of $\Delta(X)$ at the surface of the slab,
$X=L$, where $2L$ is the slab width, the effect being stronger for
smaller values of $L$. In more detail, the gap equation for the
nuclear slab was solved in \cite{Pankr1} for the Paris force and
in \cite{Pankr2} for the Argonne v$_{18}$ force. It turned out
that predictions of these two absolutely different kinds of
realistic NN-force for the gap function agree with each other
within 10\%, both yielding the ratio $\Delta(X\simeq
L)/\Delta(X=0)\simeq 2$.

Recently, Pillet et al. \cite{PiSanSch} investigated directly
spatial properties of the anomalous density $\varkappa$ which
determines the space distribution of Cooper pairs. Calculations
were carried out within the HFB approach with employing the D1S
Gogny interaction \cite{Gogny} for a set of Sn, Ni, and Ca
isotopes. It was shown that Cooper pairs in nuclei preferentially
are located with small size ($2- 3\;$fm) in the surface region.
The relevance of this phenomenon to two-nucleon transfer reactions
was discussed. It should be mentioned that earlier the correlation
properties of pairing for specific nuclei were studied by Catara
et. al. \cite{Catara}, Ferreira et al. \cite{Ferreira}, Bertsch et
al. \cite{Bertsch}, Hagino et al. \cite{a12} and yet in several
works cited in \cite{PiSanSch}. A similar investigation has also
been performed for $T=0$ pairing in dilute nuclear matter
\cite{L_Sch}.

In this paper we carry out an analogous study for a nuclear slab
with realistic NN force (the Paris and Argonne v$_{18}$
potentials) and the Gogny force. Our goal is to compare
predictions for the correlation pairing characteristics of the
Gogny force and of realistic forces in order to analyze, to what
extent the effect found in \cite{PiSanSch} is general and
independent on the specific choice of NN force. It should be
mentioned that, with small modifications, the nuclear slab
configuration may be used to describe the so-called ``lasagna''
phase of the inner crust of neutron stars.

\section{Main definitions}

To make the comparison easier, let us recall the  main definitions
introduced in \cite{PiSanSch}.  In a inhomogeneous system, the
anomalous density matrix is defined as follows: \beq
\varkappa({\bf r}_1,{\bf r}_2) = \sum_i u_i({\bf r}_1) v_i({\bf
r}_2), \label{kapuv} \eeq where $u_i({\bf r}),v_i({\bf r})$ are
the Bogolyubov functions. For a spherical nucleus, it is
convenient to go to relative and center of mass coordinates, ${\bf
r}={\bf r}_1-{\bf r}_2$ and ${\bf R}=({\bf r}_1+{\bf r}_2)/2$. In
\cite{PiSanSch} the anomalous density matrix was studied in such a
way that the probability distribution of Cooper pairs,
$|\varkappa(\bR,\br)|^2$, was averaged over the angle between the
vectors ${\bf R}$ and ${\bf r}$, \beq \varkappa^2(R,r)=\frac 1
{4\pi}\int |\varkappa(\bR,\br)|^2 \; d\Omega \;.\label{avr}\eeq In
particular, the space distribution of the pairing tensor
$|\varkappa(R,r)|^2$ was analyzed. The probability distribution of
pairing correlations, \beq P(R,r)=R^2 r^2
\varkappa^2(R,r)\,,\label{Prob} \eeq was calculated in
\cite{PiSanSch}. To avoid misunderstanding, this quantity is not
normalized to unity.

The coordinate dependent local correlation length was defined as
\beq \xi(R) = \frac{(\int
r^2\varkappa^2(R;r)\;d^3r)^{1/2}}{(\int\varkappa^2(R;r)\;d^3r)^{1/2}}\;.
\label{cor_leng} \eeq At last, the locally normalized pairing
tensor was considered in the form \beq W(R,r) = \dfrac{r^2
\varkappa^2(R,r)}{\int \varkappa^2(R,r)\; r^2 dr}\;,
\label{Wr}\eeq as it enters the definition of the correlation
length.

Just as in \cite{Pankr2}, we consider a nuclear slab embedded into
the Saxon-Woods potential $U(x)$ symmetrical with respect to the
point $x=0$ with  potential well depth $U_0=-50\;$MeV and
diffuseness parameter of $d=0.65\;$fm typical for finite nuclei.
The chemical potential is taken equal to $\mu=-8\;$MeV. To compare
our calculations with those of \cite{PiSanSch}, we fixed the
thickness parameter of the slab as $L=6\;$fm to mimic nuclei of
the tin region. We use the notation $\br=(\bs,x)$, where $\bs$ is
the two-dimensional vector in the plane perpendicular to the
 $x$-axis. The system under consideration is homogeneous
 in the $\bs$-plane, therefore one has $\varkappa({\bf r}_1,{\bf r}_2)\to \varkappa(\bR,\br)
 \to \varkappa(X;x,\bs)$, with the obvious notation. The
 definition (\ref{cor_leng}) is then rewritten as follows:
\beq \xi^2(X) =
\dfrac{\int(x^2+s^2)|\varkappa(X;x,\bs)|^2\;d^3r}{\int|\varkappa(X;x,
 \bs)|^2\;d^3r}\;. \label{cor_lr} \eeq As far as the correlation
properties in the $x$-direction and in the $\bs$-plane are
essentially different, it looks reasonable to consider them
separately, \beq \xi^2(X) =\xi_x^2(X)+\xi_s^2(X)\;,
\label{cor_lxs} \eeq with the obvious notation.

In the slab geometry, the angular averaging procedure similar to
that in Eq.~(\ref{avr}) is as follows:\beq \varkappa^2(X,r)=\frac
1 {4\pi}\int |\varkappa(X;x,\bs)|^2 \; \dfrac 2{r}\;
\delta(r^2-x^2-s^2)\;d^2s\;dx \;.\label{avr_r}\eeq It gives the
distribution of the pairing tensor for a fixed value of the
3-dimensional relative distance $r$. There is another possibility,
just to integrate over $\bs$: \beq \varkappa^2(X,x)= \int
|\varkappa(X;x,\bs)|^2 \; d^2s \;.\label{avr_x}\eeq It yields the
quantity $\varkappa^2(X,x)$ which gives the distribution of the
pairing tensor in natural for slab variables. For brevity, we use
the same notation for the integrated anomalous density matrix as
for the initial one and the one in Eq.~(\ref{avr_r}). The
arguments should help to avoid misleading. Note that
$\varkappa(X,x)$ and $\varkappa(X,r)$ have different dimensions.
\begin{figure}[]
\centerline{\includegraphics [height=80mm]{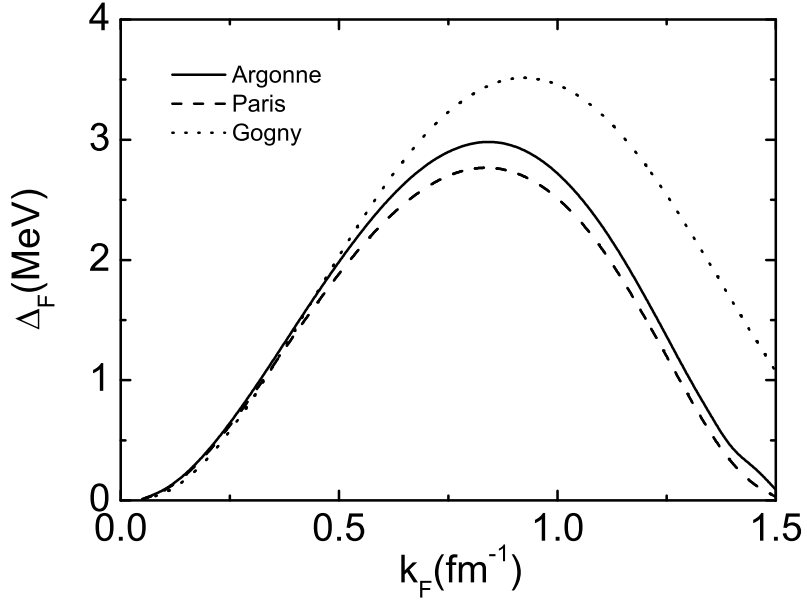}} \caption{The gap
$\Delta(k=k_{\Fs})$ in infinite nuclear matter }\label{Delt_inf}
\end{figure}

\begin{figure}[]
\centerline{\includegraphics [height=80mm]{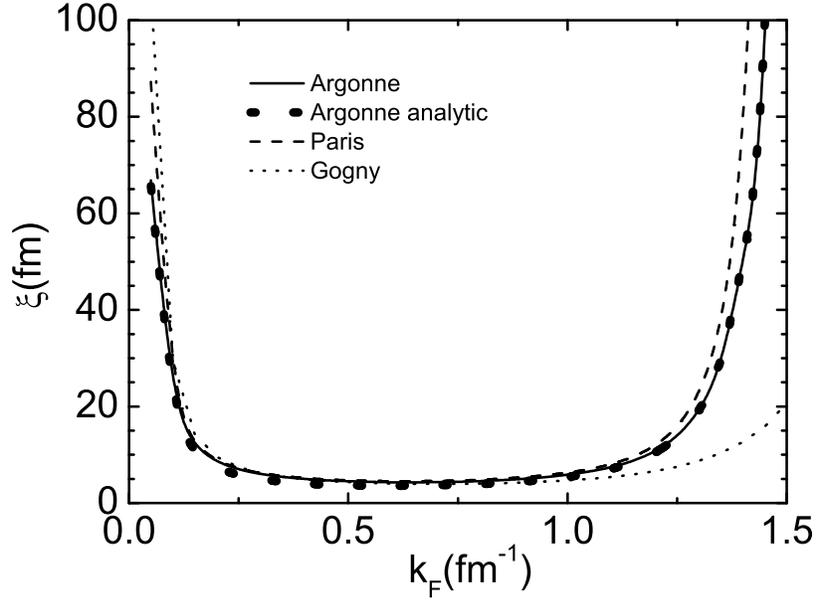}} \caption{The
correlation pairing length in infinite nuclear
matter}\label{fig_leng_inf}
\end{figure}

\section{Calculation results}

Methods of solving the gap equation and the Bogolyubov equations
for a nuclear slab are described in \cite{Pankr1} for the
separable representation of the Paris potential and in
\cite{Pankr2}, for the Argonne v$_{18}$ force. The latter could be
used for arbitrary NN-potential, and we repeated all the
calculations of \cite{Pankr2} for the Gogny force. To begin the
comparison, let us start from  infinite nuclear matter. The
dependence of the gap on the density of nuclear matter for the
three kinds of the NN-force is displayed in Fig. \ref{Delt_inf}.
The correlation length (\ref{cor_leng}) in infinite matter  can be
easily found in the momentum space: \beq \xi^2  = \dfrac {\int
\left|\dfrac{\partial}{\partial k} \varkappa(k)\right|^2 d^3k }
{\int |\varkappa(k)|^2 d^3k}. \label{cor_inf}\eeq  Let us
substitute in this equation $\varkappa(k)=\Delta(k)/2E_k$, where
$E_k=\sqrt{(\eps_k-\eps_{\Fs})^2+\Delta^2(k)}$, $\eps_k=k^2/2m^*$,
$\eps_{\Fs}=k_{\Fs}^2/2m^*$. The functions inside the integrals
both in the numerator and denominator of this relation are very
peaked in the vicinity of $k=k_{\Fs}$ and  rapidly vanish outside
the interval $|k-k_{\Fs}|\lsim k_{\Fs} (\Delta_{\Fs}/\eps_{\Fs})$,
$\Delta_{\Fs}=\Delta(k_{\Fs})$. Usually one deals with the limit
$(\Delta_{\Fs}/\eps_{\Fs})<<1$. In this case, one can substitute
$\Delta(k)=\Delta_{\Fs}$ in Eq.~(\ref{cor_inf}) and evaluate the
integrals analytically. A simple calculation yields \beq \xi  =
\frac {v_{\Fs}}{\sqrt{8}\Delta_{\Fs}}\;, \label{cor_an} \eeq where
$v_{\Fs}=k_{\Fs}/m^*$.

The correlation length for the three kinds of force under
discussion found numerically from Eq.~(\ref{cor_inf}), with
$m^*=m$, are displayed  in Fig. \ref{fig_leng_inf}. For
comparison, the approximate $\xi$ from Eq.~(\ref{cor_an}) with
Argonne force is also displayed.  It is seen that the approximate
formula works sufficiently well in all the interval of $k_{\Fs}$.
Even at the maximum of the gap the deviation from the numerical
result is of the order of 15\%.

One can see that at small density, $k_{\Fs}<0.5\;$fm$^{-1}$, the
results for all three forces practically coincide. This is not
strange. Indeed, although the Gogny force is an effective one, in
the $^1S_0$ channel under consideration it describes the free
NN-scattering perfectly well for small energy values which are
only important in this density interval. In this sense, the Gogny
force could be considered as a semi-realistic force. Two realistic
forces lead to close results for all density values. In the
vicinity of the gap maximum, the difference between $\Delta^{\rm
Arg}$ and $\Delta^{\rm Par}$ does not exceed 10\%, and only at
$k_{\Fs}\simeq 1.4 \;$fm$^{-1}$, where the gap value itself
becomes very small, the relative difference becomes larger. As to
the Gogny force, at the density region $k_{\Fs}\simeq 1
\;$fm$^{-1}$ it leads to the gap values which are bigger by
approximately $(20\div 30)$\% than those for realistic forces.
Correspondingly, the correlation length for the Gogny force is
quite close to that of the realistic forces till $k_{\Fs}\simeq
0.8 \;$fm$^{-1}$, and only at $k_{\Fs}\simeq 1.2 \;$fm$^{-1}$ the
difference becomes large. The density dependence of the
correlation length, $\xi(k_{\Fs})$, is qualitatively similar for
all the three types of force. It consists of a plateau at
$0.3\lsim k_{\Fs}\lsim 1 \;$fm$^{-1}$ and two intervals of sharp
growth, at $ k_{\Fs}\lsim 0.3 \;$fm$^{-1}$ and $ k_{\Fs}\gsim 1
\;$fm$^{-1}$. In the latter, the value of $\xi^{\rm Gog}(k_{\Fs})$
is growing with $k_{\Fs}$ much slower than that of $\xi^{\rm
Arg}(k_{\Fs})$ and $\xi^{\rm Par}(k_{\Fs})$. Note that at $
k_{\Fs}\gsim 1.2 \;$fm$^{-1}$ the difference between $\xi^{\rm
Arg}(k_{\Fs})$ and $\xi^{\rm Par}(k_{\Fs})$ also becomes
noticeable. This is a manifestation of their behavior near the
critical point $ k_{\Fs}^{\rm c}$ at which the gap vanishes and
transition to the normal phase of nuclear matter occurs. The
values of $ k_{\Fs}^{\rm c}$  for the Argonne force and the Paris
potential are a little different. This results in different
behavior of $\xi^{\rm Arg}(k_{\Fs})$ and $\xi^{\rm Par}(k_{\Fs})$
in the region of $ k_{\Fs}\simeq 1.5 \;$fm$^{-1}$.

\begin{figure}[]
\centerline{\includegraphics [height=80mm]{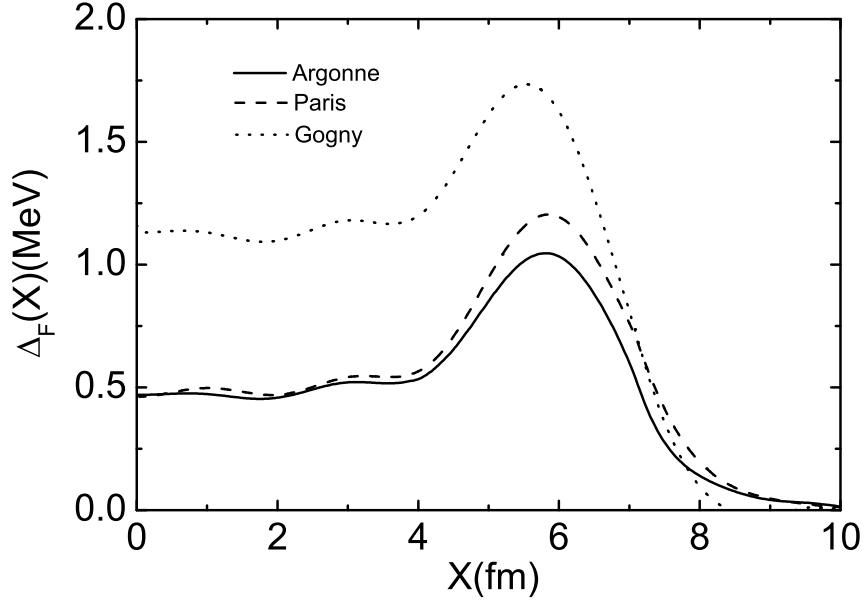}} \caption{The
Fermi averaged gap $\Delta_{\Fs}(X)$}\label{D_F}
\end{figure}

\begin{figure}[]
\centerline{\includegraphics [height=80mm]{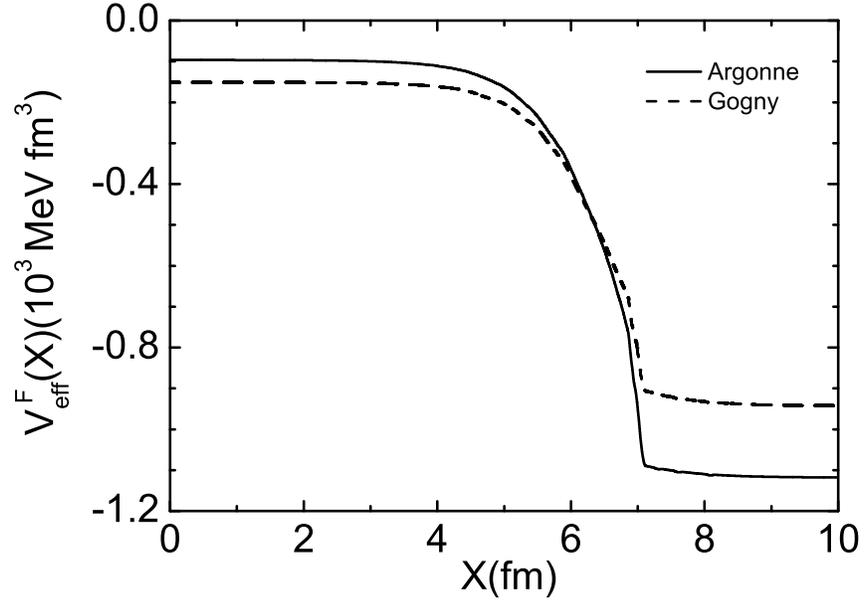}} \caption{The
Fermi averaged effective pairing interaction $\VF (X)$}
\label{Fig_Vef}
\end{figure}

\begin{figure}[]
\centerline{\includegraphics [height=80mm]{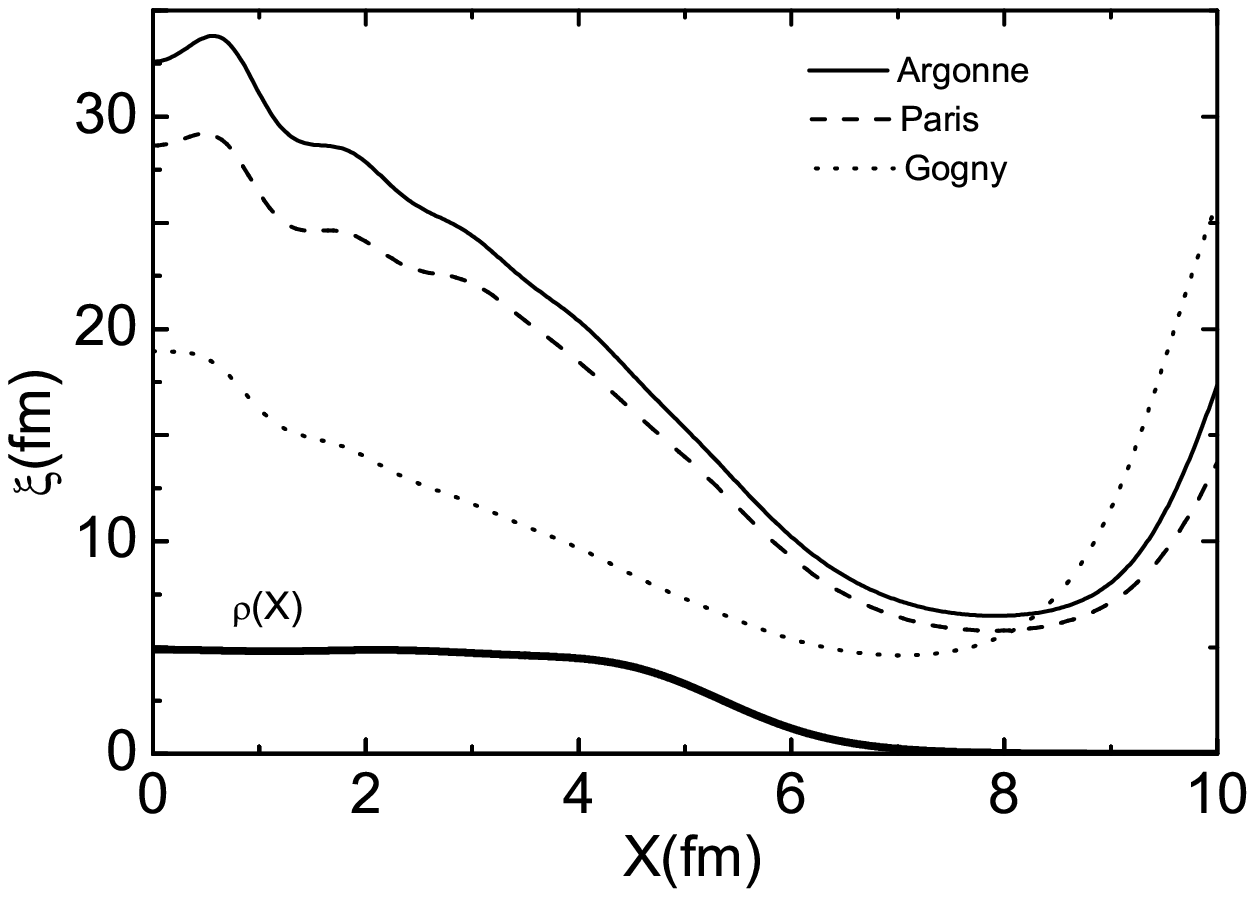}} \caption{The
correlation pairing length $\xi(X)$ in a slab of nuclear
matter}\label{leng_r}
\end{figure}

\begin{figure}[]
\centerline{\includegraphics [height=80mm]{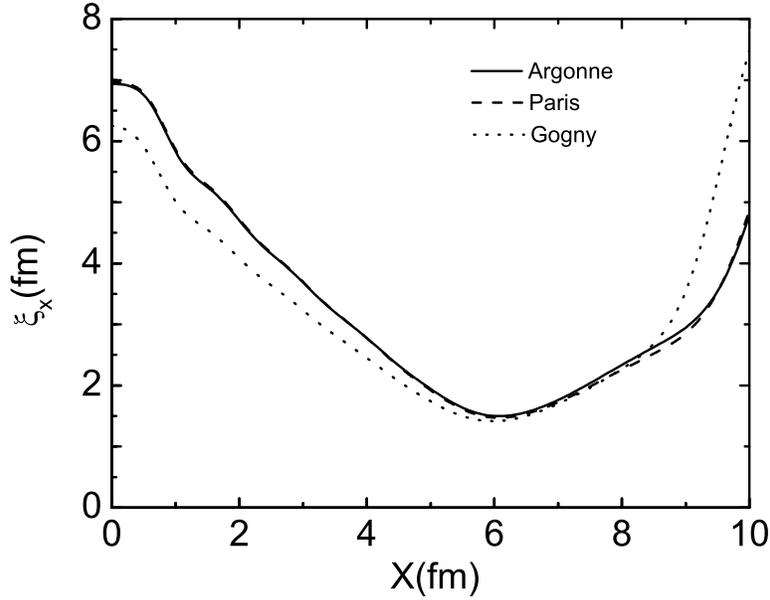}} \caption{The
correlation pairing length $\xi_x(X)$ in the $x$-direction for a
slab of nuclear matter}\label{leng_x}
\end{figure}

 Let us now turn to the slab system. Before analyzing  the
 correlation characteristics, it is instructive to briefly compare the
 EPI and the gap itself found with the realistic forces and the
 Gogny force. The ``Fermi averaged'' gap is displayed in Fig.\ref{D_F}.
 It is defined as
follows: \beq \Delta_{\Fs}(X) = \Delta(X,k_{\Fs}(X))\,,
 \label{Del_F} \eeq where the local Fermi momentum is
$k_{\Fs}(X){=}\sqrt{2m(\mu{-}U(X))}\;\Theta(\mu{-}U(X))$. This
quantity characterizes the gap on average
\cite{Rep,Pankr1,Pankr2}. We see, first, that all three functions
$\Delta_{\Fs}(X)$ have pronounced maxima at $X\simeq L=6\;fm$. The
ratio  $\Delta_{\Fs}(X\simeq L)/\Delta_{\Fs}(X=0) \simeq 2$ for
realistic forces and $\simeq 1.5$ for the Gogny force, in
agreement with \cite{6auth}. Second, the gap $\Delta_{\Fs}^{\rm
Gog}$ is significantly bigger than $\Delta_{\Fs}^{\rm Par}$ and
$\Delta_{\Fs}^{\rm Arg}$, by approximately a factor one and a half
at the surface and two  inside the slab. It is worth to discuss
this point in more detail. Let us  consider $^{120}$Sn as a
``reference nucleus''.  Its empirical gap value is estimated
usually as $\Delta\simeq 1.3\;$MeV \cite{Fay,milan}. Diagonal
matrix elements of the gap found in \cite{Pankr1,Pankr2} for a
slab with $L=6\;$fm are about 1 MeV which agrees with the above
value, leaving a room about 20--30\% for the surface vibration
contribution. The latter was estimated in \cite{milan} as $\simeq
50$\% which is, in our opinion, too much, see discussion in
\cite{Rep}. We consider the estimation of
 \cite{A_Kam} at $\simeq 30$\%  as more
realistic, but, evidently, also too big, because of disregarding
so-called tadpole diagrams \cite{Kam_S}. Calculations of
\cite{PiSanSch} for this nucleus gave $\Delta\simeq 2\;$MeV which,
in our opinion, is too much, especially if to take into account
that some additional contribution of surface vibrations to the
mean field theory value of $\Delta$ should be! Thus, our
observation in slab calculations that the Gogny force
overestimates the gap value agrees essentially with results of
\cite{PiSanSch}.

To understand the physical reason for the surface enhancement of
the pairing gap with each NN force under consideration and bigger
values of the gap for Gogny force, it is useful to calculate the
EPI which we use in the two-step method of solving the gap
equation \cite{Rep}. Let us remind how this quantity is defined.
In a symbolic form, the microscopic gap equation reads: \beq\Delta
= {\cal V} A^{s}\Delta\,,\label{del}\eeq where ${\cal V}$ is the
free $NN$-potential and $A^s=GG^s$ stands for the two-particle
propagator in the superfluid system. Here $G$ and $G^s$ are the
one-particle Green functions without and with pairing effects
taken into account, respectively. In Eq.~(\ref{del}), as usual,
integration over intermediate coordinates and summation over spin
variables is understood. Let us now split the complete Hilbert
space $S$ of two-particle states into two parts, $S=S_0+S'$. The
first one is the model subspace $S_0$ in which the gap equation is
considered, and the other is the complementary subspace $S'$. They
are separated by the energy $E_0$ in such a way that $S_0$
involves all the two-particle states $(\lambda ,\lambda')$ with
the single-particle energies $\eps_{\lambda}, \eps_{\lambda'}
<E_0$. The complementary subspace $S'$ involves the two-particle
states for which one of the energies
$\eps_{\lambda},\eps_{\lambda'}$ or both of them are greater than
$E_0$. Therefore, pairing effects can be neglected in $S'$ if
$E_0$ is sufficiently large. The validity of inequality
$\Delta^2/(E_0-\mu)^2\ll 1$ is the criterium of such
approximation. Correspondingly, the two-particle propagator is
represented as the sum $A^s = A^s_0 + A'$. Here we already
neglected  the superfluid effects in the $S'$-subspace and omitted
the superscript ``s'' in the second term. The gap equation
(\ref{del}) can be rewritten in the model subspace, \beq \Delta =
\Vef\,A^{s}_{0}\,\Delta\,,\label{del0}\eeq where the EPI should be
found in the supplementary subspace, \beq\Vef = {\cal V} + {\cal
V} A'\,\Vef\,.\label{EPI}\eeq Note that the last equation has a
strong similarity with the Bethe--Goldstone equation.

As the analysis showed \cite{Pankr1,Pankr2}, the optimal choice of
splitting corresponds to $E_0=15\div20\;$MeV. In a slab system,
the EPI is calculated in the mixed coordinate-momentum
representation \cite{Rep}. To illustrate graphically properties of
the EPI, we present it in a localized form \cite{Pankr1,Pankr2}
with the Fermi averaged strength \beq \VF (X) = \int dt \Vef
(k_1{=}k_2{=}k_{\Fs}(X);X+\frac t{2},X-\frac t{2} )\,.
 \label{VF} \eeq
 The Fermi averaged EPI for the
Argonne v$_{18}$ force and the Gogny D1S force calculated for
$E_0=15\;$MeV  are displayed in Fig.\ref{Fig_Vef}. We did not
display the EPI for the Paris potential as it practically
coincides with that for the Argonne force. One can see that both
the curves behave in a similar way changing from quite week
attraction inside the slab to very strong one outside. The reason
for the latter is that in the asymptotic region $X>L$ the  $\VF$
value tends to the quantity which is very close to the free
$T$-matrix taken at the negative energy $E=2\mu$. To be precise,
the limit is equal to $T(E=2\mu)$ if the separating energy
$E_0=0$. In the case of $E_0 \ne 0$ the limit is equal to some
``$T'$-matrix'' which is obtained by solving the same
Lippman--Schwinger equation as the $T$-matrix, but in a cut
momentum space, because the contribution of nucleons with total
energy less than $E_0$ must be pulled out. As it is known, the
Gogny force leads to the scattering length in the $^1S_0$ channel
$a\simeq 12\;$fm \cite{Bertsch,a12}. It differs, of course, from
the experimental value of $a\simeq 18\;$fm which is reproduced by
any realistic force, but not so much. In any case, the virtual
pole of the $T$-matrix for the Gogny force is close to zero as it
should be. Therefore the analytical continuation of the $T$-matrix
(or $T'$-matrix) to rather small negative energy $E=2\mu$ results
in an enhancement of the $T'$-matrix ($\simeq -950\;$MeV fm$^3$)
in comparison with typical values. This enhancement is not so
strong as in the case of the Argonne force ($\simeq -1100\;$MeV
fm$^3$), but it is equally significant. The inner value of the EPI
for the
 Gogny force is quite small ($\simeq -160\;$MeV fm$^3$), but bigger
than for the Argonne one ($\simeq -95\;$MeV fm$^3$). In our
previous study with realistic forces \cite{Rep,Pankr1,Pankr2} we
explained the surface enhancement of the gap in terms of the sharp
variation of the EPI at the surface. In the case of the Argonne
force, the ratio ${\cal V}^{\rm out}_{\mbox{\scriptsize
eff}}/{\cal V}^{\rm in}_{\mbox{\scriptsize eff}}\simeq 12$. For
the Gogny force, it is about 6. This is also a big number leading
to a surface enhancement of the gap, but not so pronounced as for
realistic force. Fig.\ref{Fig_Vef} explains also why the Gogny gap
is so big. It is well known that the pairing gap depends on the
interaction strength in an exponential way. In \cite{Pankr1} it
was found that 1\% variation of ${\cal V}^{\rm
in}_{\mbox{\scriptsize eff}}$ leads to 5\% variation of the gap.
It explains why the gap function for the Gogny force is in
$1.5\div2$ times greater than the one for realistic forces.

\begin{figure}[]
\centerline{\includegraphics [height=120mm]{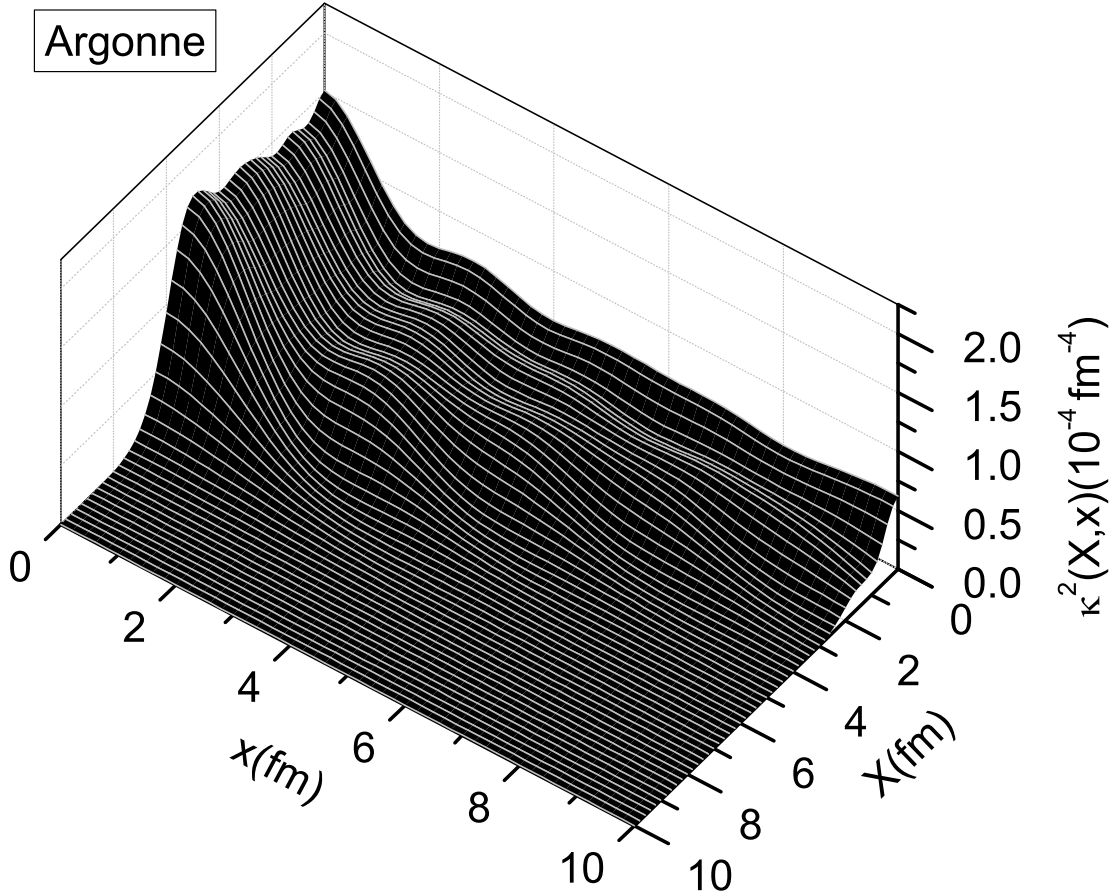}} \caption{The
pairing tensor distribution $\varkappa^2(X,x)$ }\label{kap_sq}
\end{figure}

\begin{figure}[]
\centerline{\includegraphics [height=120mm]{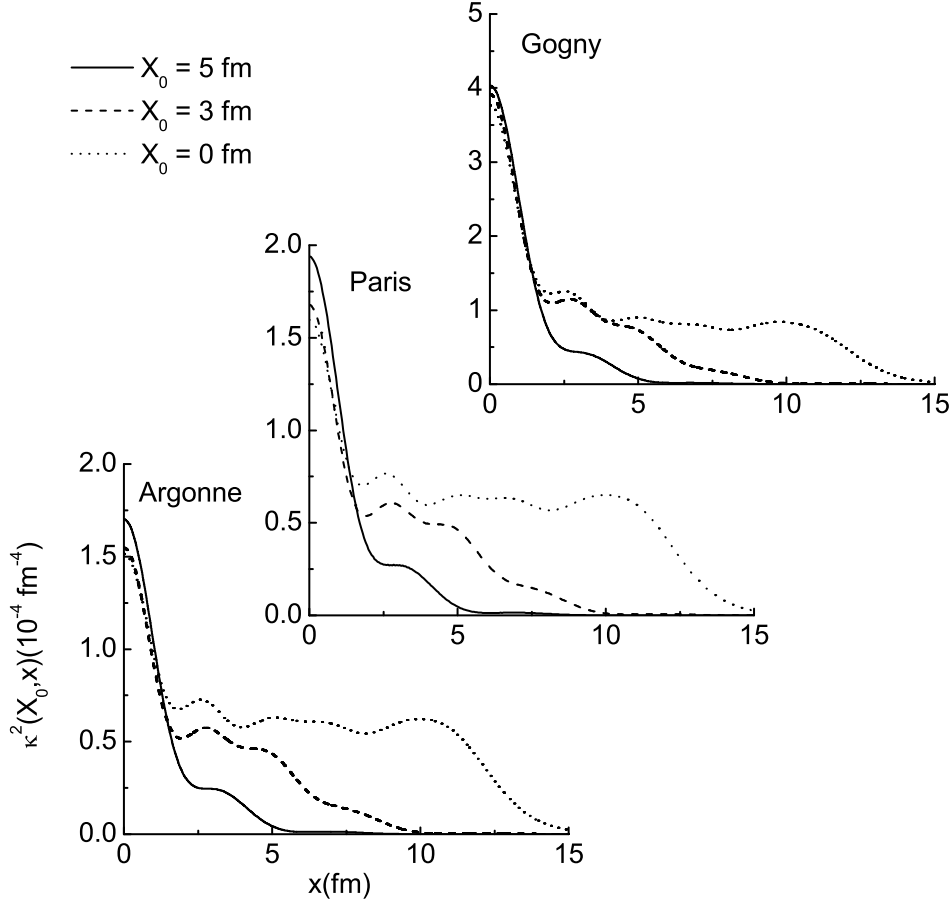}} \caption{The
profile functions $\varkappa^2(X=X_0,x)$}\label{prof-x}
\end{figure}

\begin{figure}[]
\centerline{\includegraphics [height=120mm]{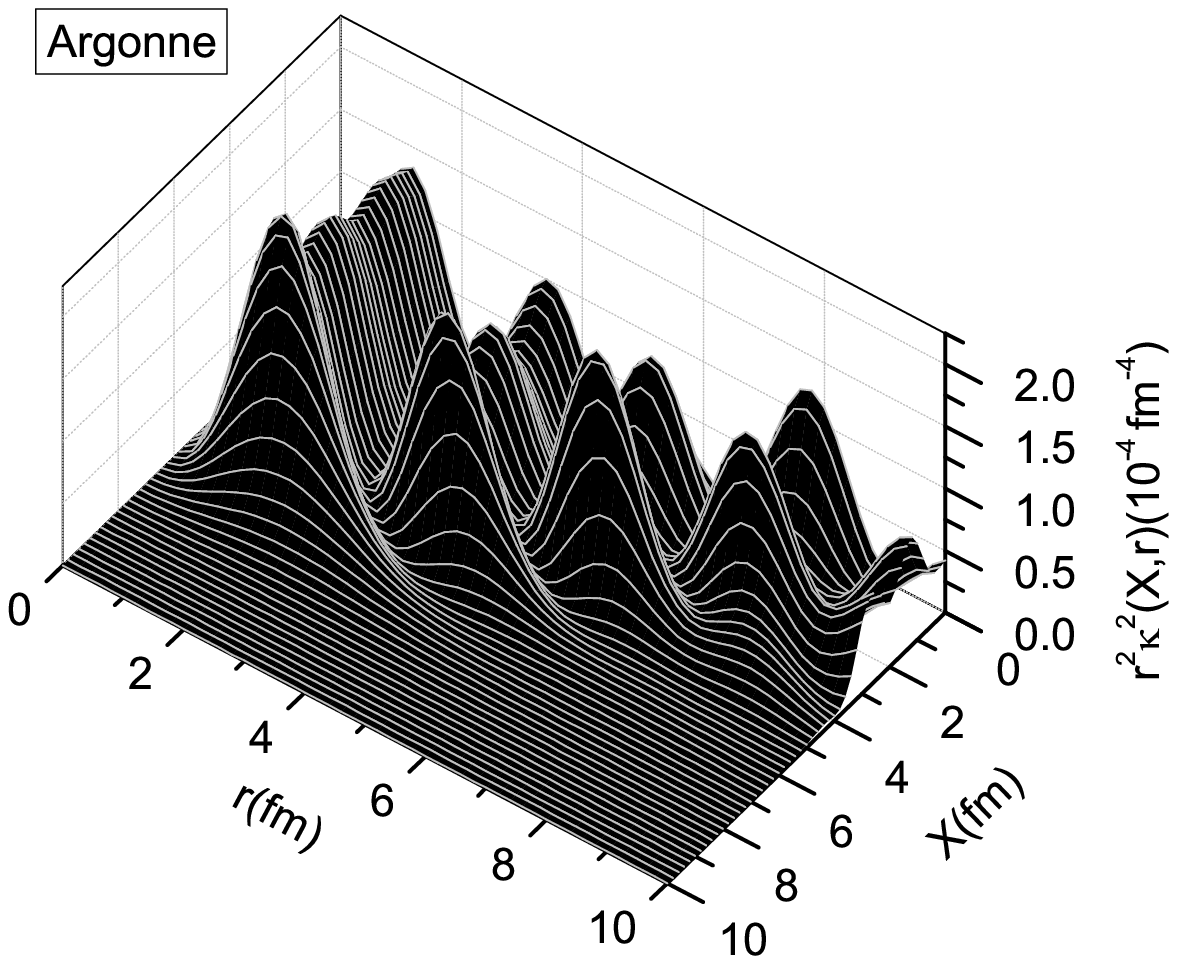}} \caption{The
probability distribution $r^2\varkappa^2(X,r)$ }\label{Prob_r}
\end{figure}

\begin{figure}[]
\centerline{\includegraphics [height=120mm]{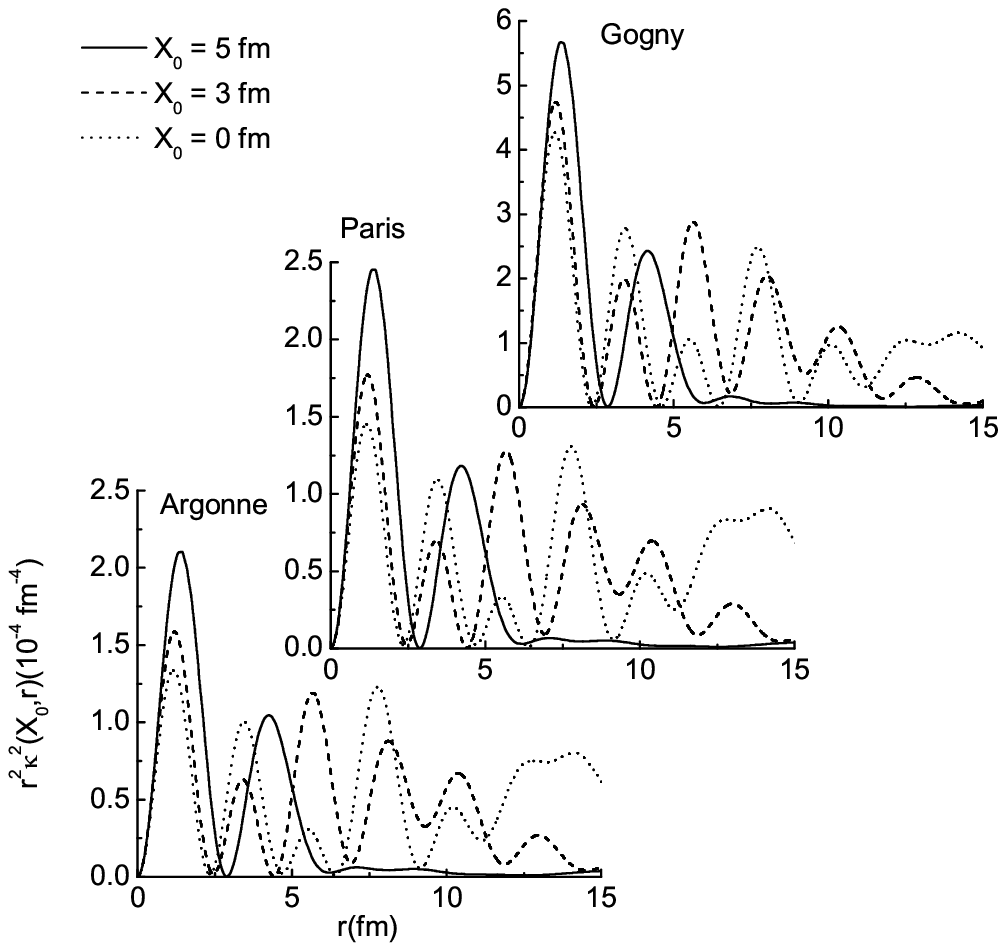}} \caption{The
profile functions $r^2\varkappa^2(X=X_0,r)$}\label{prof-r}
\end{figure}

\begin{figure}[]
\centerline{\includegraphics [height=120mm]{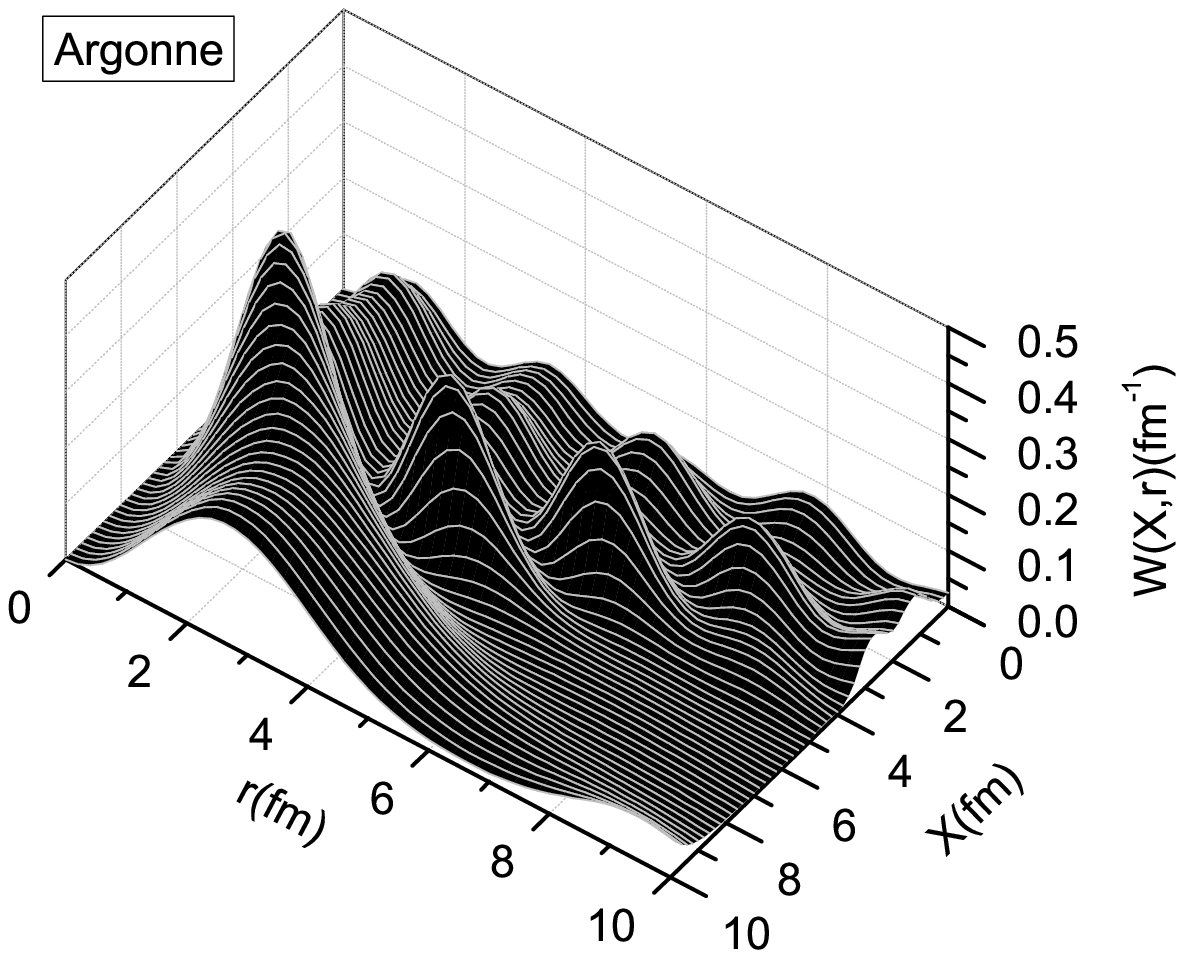}} \caption{The
locally normalized pairing tensor $W(X,r)$ }\label{fig_W_r}
\end{figure}

Let us turn now to the correlation pairing characteristics. The
total correlation lengths $\xi$ and the correlation lengths in
$x$-direction $\xi_x$ found from Eq.~(\ref{cor_lxs}) for each kind
of force are displayed in
 Fig.~\ref{leng_r} and  Fig.~\ref{leng_x}, correspondingly.
 The curve $\xi^{\rm Arg}(X)$ is quite similar to $\xi^{\rm
 Par}(X)$, the difference is about 5\%. Both have a minimum
 shifted a little from the surface, $X=L$, in direction to the
 free space, both have a pronounced maximum in the vicinity of $X=0$
 and grow rapidly to the right from the minimum. Qualitatively,
 the curve $\xi^{\rm Gog}(X)$ behaves in a similar way, but the maximum value at
$X=0$ is approximately two times less than those for realistic
forces. The minimum of $\xi^{\rm Gog}(X)$ is also shifted from the
point $X=L$, but the value of the shift is less. Such a behavior
of each curve $\xi(X)$ qualitatively agrees with naive local
density approximation (LDA) predictions. Indeed, inside the slab,
the local Fermi momentum $k_{\Fs}(X\simeq 0) \simeq
\sqrt{2m(\mu-U_0)} \simeq 1.4\;$fm$^{-1}$ which corresponds to big
values of $\xi$ in infinite matter (see Fig.~\ref{fig_leng_inf}).
The same is true at $X>8\;$fm where the local Fermi momentum
vanishes. But under more detailed examination, deviations from the
LDA predictions are significant. To illustrate this point, we
display in Fig.~\ref{leng_r} with a thick line, in arbitrary
units, the density distribution $\rho(X)$. Within the LDA, the
correlation length $\xi(X)$ should show a plateau inside the slab,
just as the density does. Instead, $\xi(X)$ is decreasing rapidly
with increase of $X$ till $X\simeq L$. Qualitatively, the
coordinate dependence of the function $\xi(X)$ in the slab reminds
us that of $\xi(R)$ in  spherical nuclei found in \cite{PiSanSch}.
Both of them have minima at the surface region and pronounced
maxima in the center. But any quantitative comparison is hardly
possible since, for the problem under consideration, the
properties of the two systems are essentially different. Indeed,
in a spherical nucleus all particles move in a finite space
limited by the nuclear surface. On the contrary, in a slab the
particle motion is limited only in the $x$-direction. In the ${\bf
s}$-plane, the motion is free which leads to very big values of
$\xi_s$ in Eq.~(\ref{cor_lxs}) and $\xi$ close to that in infinite
system. As to the $\xi_x(X)$ function, it should be much closer to
$\xi(R)$ of \cite{PiSanSch}. As it is seen in Fig.~\ref{leng_x},
two curves $\xi^{\rm Arg}_x(X)$ and $\xi^{\rm
 Par}_x(X)$ practically coincide. Deviation of  $\xi^{\rm Gog}_x(X)$
from the both is much less than in the case of the total
correlation length. It doesn't exceed 15\%. All three curves have
common minimum at $X\simeq L$, with the value of
$\xi_x^{\min}\simeq 1.5\;$fm. It is not far from the value of
$\xi^{\min}\simeq 2\;$fm found in \cite{PiSanSch} for Sn isotopes.
Evidently, the difference is mainly due to geometry effects. Thus,
for a nuclear slab, the correlation length of pairing in the
$x$-direction at the surface, calculated with realistic and
semi-realistic Gogny forces, is very small, in agreement with
conclusions of \cite{PiSanSch}.

To visualize the pairing tensor distribution, we draw a
3-dimensional plot in Fig.\ref{kap_sq} for the $\varkappa^2(X,x)$
function given by Eq.~(\ref{avr_x}) for the Argonne force. We see
that there is a set of maxima at $x=0$, the highest one being near
to the slab surface, $X\simeq L$. Fig.\ref{prof-x} shows the
profile functions $\varkappa^2(X=X_0,x)$ for several values of
$X_0$ corresponding to the maximum positions. The nearest to the
surface maximum there is at $X_0\simeq 5\;$fm, the neighboring
one, at $X_0\simeq 3\;$fm. There is also a pronounced maximum at
$X_0=0$. The surface peak is very narrow, in correspondence with
Fig.\ref{leng_x}. On the contrary, in the case of $X_0=0$ a
comparatively sharp peak is accompanied by a flat base plate which
results in a big value of $\xi_x$. The similar profile functions
are drawn, in the same figure, for the Paris and Gogny force. The
Paris curves are again quite similar to the Argonne ones. As to
the Gogny force, the similarity takes place only at qualitative
level, absolute values of $\varkappa^2$ being bigger than those
for realistic forces. This is the result of bigger values of the
gap itself for the Gogny force, as illustrated in Fig. \ref{D_F}.
We see that $\Delta_{\Fs}^{\rm Par}$ is a little bigger than
$\Delta_{\Fs}^{\rm Arg}$, therefore $(\varkappa^{\rm Par})^2$ is,
on average, bigger than $(\varkappa^{\rm Arg})^2$. On the other
side, $\Delta_{\Fs}^{\rm Gog}$ is significantly bigger than the
gap for realistic forces. As the result, $(\varkappa^{\rm Gog})^2$
 significantly exceeds the microscopic values as well. Note that the correlation
length $\xi(X)$, Eq.~(\ref{cor_lr}), does not depend on the
magnitude of $\varkappa$, giving rise to a much smaller difference
between the Gogny and realistic forces. The same is true for
$\xi_x(X)$, Eq.~(\ref{cor_lxs}). The surface peak dominates, to
some extent, for all three kinds of forces, and this effect for
realistic forces is even stronger than for Gogny force.

The $\varkappa^2(X,r)$ function given by Eq.~(\ref{avr_r})
multiplied by $r^2$ gives the probability distribution of paring
correlations similar to Eq.~(\ref{Prob}) in the case of the
spherical symmetry. It is displayed in Fig.\ref{Prob_r}, again for
the Argonne force. Now the main maximum positions are shifted from
$r=0$ to $r\simeq 1 \div 2\;$fm due to the factor $r^2$. The
surface maximum at $X\simeq L$ is even more pronounced  than in
Fig.\ref{kap_sq}. To compare results obtained for different forces
under consideration, we again draw the profile functions
$r^2\varkappa^2(X=X_0,r)$, see Fig.\ref{prof-r}. Again, just as in
Fig.\ref{prof-x},  the  Gogny force results are significantly
bigger than those  of realistic ones in the magnitude but are very
similar in the form. Again the surface maxima dominates and again
the surface enhancement is  stronger for realistic forces.

\begin{figure}[]
\centerline{\includegraphics [height=80mm]{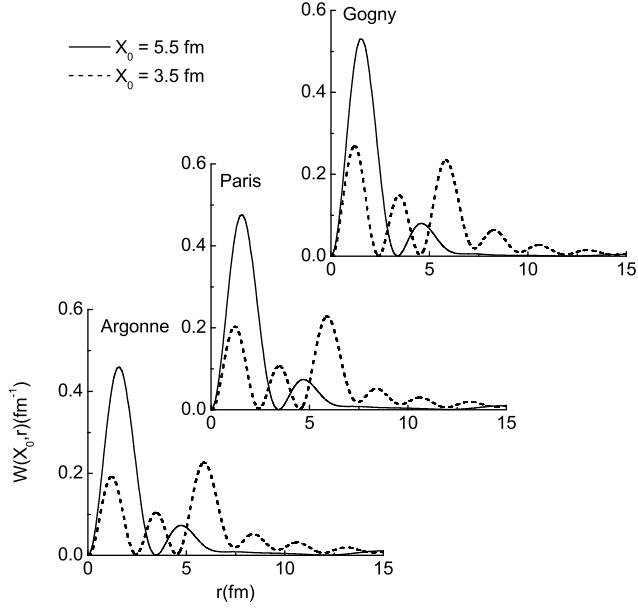}} \caption{The
profile functions $W(X_0,r)$}\label{profileW}
\end{figure}

To make the comparison with \cite{PiSanSch} more complete, we
display in Fig.\ref{fig_W_r} the locally normalized pairing tensor
which is defined as \beq W(X,r) =
\dfrac{r^2\varkappa^2(X,r)}{\int\varkappa^2(X,r)\;d^3r}\;,
\label{W_r} \eeq similar to the definition (\ref{Wr}) in spherical
systems. The profile functions $W(X=X_0,r)$ are displayed in
Fig.\ref{profileW}
 which is analogous to Fig.\ref{prof-x} and Fig.\ref{prof-r}.
 In this case, the Gogny force curves are close to those for
 realistic forces not only in the form but also in the magnitude.
 This is happens because the $\varkappa^2$ quantity comes in both
 the numerator and the denominator of Eq.~(\ref{W_r}), the result
 being almost independent of the magnitude of $\varkappa$.
 Absolutely similar situation occurs with calculations of the
 correlation lengths $\xi$ and $\xi_x$.

\begin{figure}[]
\centerline{\includegraphics [height=80mm]{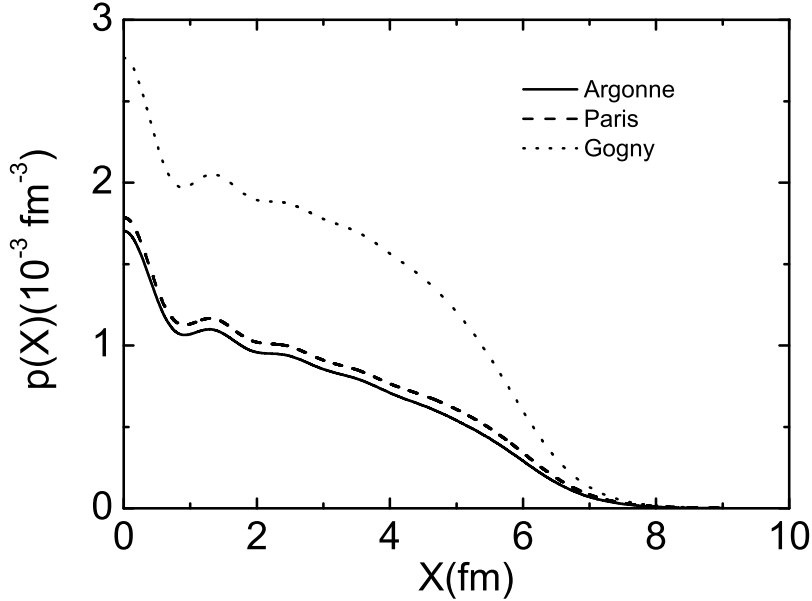}} \caption{The
integrated Cooper pair probability distribution $p(X)$
}\label{fig_p_X}
\end{figure}

The denominator of Eq.~(\ref{W_r}), \beq p(X) =
\int\varkappa^2(X,r)\;d^3r=\int\varkappa^2(X,x)\;dx\;, \label{p_X}
\eeq gives the total probability distribution of Cooper pairs
integrated over relative coordinates. Note that this quantity
displayed in Fig.\ref{fig_p_X}, just as $P(R,r)$,
Eq.~(\ref{Prob}), is not normalized to unity.   Again we see that
the result for the Gogny force behaves qualitatively similar to
those for the realistic forces but its magnitude is significantly
bigger. And for  each force under consideration this quantity in
the slab does not exhibit any surface enhancement. It occurred
because, although the surface maximum for any force in
Fig.\ref{prof-x} is higher than the central one, the correlation
length in the $x$-direction at the surface is much smaller than
inside the slab that makes the integral over $x$ to be smaller. An
analogous effect should take place in nuclei, too, but in this
case  the ``geometrical'' factor $R^2$ in Eq.~(\ref{Prob}) would
help to survive to the surface enhancement in the Cooper pairs
distribution if the probability $P(R,r)$, Eq.~(\ref{Prob}), is
integrated over the relative coordinates.

\section{Conclusion}

Spatial correlation properties of nuclear pairing in a nuclear
slab are studied with two realistic NN potentials, the Paris and
Argonne v$_{18}$, and with the phenomenological Gogny D1S
interaction. The results obtained with the two realistic forces
agree with each other with an accuracy of about 10\%. But they
agree only qualitatively with those of the Gogny force. The gap
value in the slab for the Gogny force exceeds that from the
realistic forces by a factor about 2 , which results in rather
bigger values of
 the anomalous density matrix $\varkappa$. Nevertheless, some of
main conclusions of \cite{PiSanSch} obtained for finite nuclei
with the Gogny force  are confirmed qualitatively, especially the
dependence of the correlation length on the position of the c. m.
of a Cooper pair. This quantity does not depend practically on the
absolute value of $\varkappa$, only the space distribution of the
pairing tensor $\varkappa^2(X,x)$ being important. At the surface
of the slab the local value of the correlation length in the
$x$-direction is very small, $\xi_x(X\simeq L) \simeq 1.5\;$fm,
for all three kinds of the force under consideration. Inside the
slab $\xi_x$ becomes very large, i.e. of the order of the slab
width or even more.  Thus, in this point our results completely
confirm those of \cite{PiSanSch}.

The pairing tensor $\varkappa^2(X,x=0)$ has several maxima, among
them the ones at $X=0$ and $X\simeq L$ are most pronounced. And
the one at the surface is a bit higher, especially for realistic
forces. In this sense, we can speak of  surface enhancement of the
Cooper pair distribution. However, the total probability $p(X)$
for a pair to have the c.m. coordinate $X$, which is obtained by
integrating $\varkappa^2(X,x)$ over relative coordinate, has no
surface enhancement. Thus, the second conclusion of
\cite{PiSanSch} that Cooper pairs in nuclei prefer to be
concentrated in the vicinity of the surface should not be drawn
for a slab. We explain this with different geometrical properties
of two systems under comparison, with different
``surface-to-volume ratio'' in a sphere and in a slab. However,
all surface enhancement features found for realistic forces are
qualitatively reproduced with the Gogny force. We trace this
effect to the ``semi-realistic'' nature of the Gogny force which
describes the low-energy NN-scattering in the $^1S_0$-channel
sufficiently well. It seems reasonable to suppose that all main
conclusions of \cite{PiSanSch} will be confirmed qualitatively if
the Gogny force in the pairing channel will be replaced by a
realistic NN-potential.

This research was partially supported by the Grant NSh-3003.2008.2
of the Russian Ministry for Science and Education and by the RFBR
grants 06-02-17171-a and 07-02-00553-a. Two of us (S.P. and E.S.)
thank the INFN, Seczione di Catania, for hospitality.

{}

\end{document}